\documentclass[3p,11pt]{elsarticle}
\usepackage{graphicx}

\usepackage{longtable, booktabs}
\usepackage{ltcaption}

\usepackage{amsfonts,amsmath,amssymb}
\usepackage{scalefnt}
\usepackage{pdflscape}
\usepackage[multiple]{footmisc}

\usepackage{float}  %For the Float option [H]

\newcommand{\bbOne}{1\hspace*{-0.8ex}1}
\newtheorem{thm}{Theorem}

\newtheorem{remark}{Remark}
\newtheorem{proposition}[thm]{Proposition}
\newproof{pf}{Proof}
\newtheorem{ass}{Assumption}
\newtheorem{defn}{Definition}

\journal{ }

    \makeatletter
    \def\ps@pprintTitle{%
       \let\@oddhead\@empty
       \let\@evenhead\@empty
       \def\@oddfoot{\reset@font\hfil\thepage\hfil}
       \let\@evenfoot\@oddfoot
    }
    \makeatother
    
\begin{document}

\begin{frontmatter}



\title{Pricing double barrier options on homogeneous diffusions: a Neumann series of Bessel functions representation}


\author[label1]{Igor V. Kravchenko\corref{cor1}}
\ead{ivkoh@iscte.pt}
\cortext[cor1]{Corresponding author. Tel.: +351 217650531.}

\author[label3]{Vladislav V. Kravchenko}
\ead{vkravchenko@math.cinvestav.edu.mx}

\author[label3]{Sergii M. Torba}
\ead{storba@math.cinvestav.edu.mx}

\author[label1,label2]{Jos\'{e} Carlos Dias}
\ead{jose.carlos.dias@iscte.pt}

 \address[label1]{Instituto Universit\'{a}rio de Lisboa (ISCTE-IUL), Edif\'{\i}cio II, Av. Prof. An\'{\i}bal Bettencourt, 1600-189 Lisboa, Portugal.}
 \address[label3]{Departamento de Matem\'{a}ticas, CINVESTAV del IPN, Unidad Quer\'{e}taro, Libramiento Norponiente No. 2000, Fracc. Real de Juriquilla, Quer\'{e}taro, Qro. C.P. 76230 M\'{e}xico}
 \address[label2]{Unidade de Investiga\c c\~{a}o em Desenvolvimento Empresarial (UNIDE-IUL), Lisboa, Portugal.}

\begin{abstract}
This paper develops a novel analytically tractable Neumann series of Bessel functions representation for pricing (and hedging) European-style double barrier knock-out options, which can be applied to the whole class of one-dimensional time-homogeneous diffusions even for the cases where the corresponding transition density is not known. The proposed numerical method is shown to be efficient and simple to implement. To illustrate the flexibility and computational power of the algorithm, we develop an extended jump to default model that is able to capture several empirical regularities commonly observed in the literature.
\end{abstract}


\end{frontmatter}


%\section{}
%\label{}

\section{Introduction\label{SectionIntroduction}}

In this paper we develop a novel option pricing methodology based on an analytically tractable Neumann series of Bessel functions (hereafter, NSBF) representation. The new representation is derived by applying the NSBF expansion to the arising Sturm-Liouville problem. To highlight the potential of our method, we derive a new analytical tractable representation of the price (and Greeks) of double barrier European-style knock-out options (henceforth, DBKO options), though applications to other similar problems may be designed using this conceptual framework.

Barrier option contracts are path-dependent exotic options traded at over-the-counter markets on several underlying assets, e.g., stocks, stock indexes, currencies, commodities, and interest rates. They have been actively traded mainly because they are cheaper than the corresponding vanilla options and offer an important tool for risk managers and traders to better express their market views without paying for outcomes that they may find unlikely. Moreover, they are also used as building blocks of many structured products.

Given their popularity in the market, a vast literature dedicated to their valuation has been developed. For instance, alternative pricing (and hedging) schemes for DBKO options under the classic \emph{geometric Brownian motion} (hereafter, GBM) assumption have been proposed by \cite{KunitomoIkeda1992}, \cite{GemanYor1996}, \cite{Sidenius1998}, \cite{Pelsser2000}, \cite{Schroder2000}, or \cite{BuchenKonstandatos2009}. Such modeling framework assumes the volatility is constant throughout the option's life, several attempts have been made to overcome this unrealistic assumption implicit in the GBM diffusion.

It is well-known that the \emph{constant elasticity of variance} (hereafter, CEV) diffusion model of \cite{Cox1975}, where the volatility is a function of the underlying asset price, is able to better reproduce two empirical regularities commonly observed in the literature, namely the existence of a negative correlation between stock returns and realized volatility (\emph{leverage effect}) and the inverse relation between the implied volatility and the strike price of an option contract (\emph{implied volatility skew}). To accommodate these observations, the valuation of DBKO options under the CEV model have been performed by \cite{BoyleTian1999} through a trinomial scheme, by \cite{DavydovLinetsky2001} using a pricing framework based on the numerical inversion of Laplace transforms, by \cite{DavydovLinetsky2003} via an eigenfunction expansion approach, and by \cite{MijatovicPistorius2013} whose approach rests on the construction of an approximation based on continuous-time Markov chains, amongst others.

More recently, \cite{DiasNunesRuas2015} tackle the valuation of DBKO options (using the stopping time approach as well as the static hedging approach) under the so-called \emph{jump to default CEV} (hereafter, JDCEV) model proposed by \cite{CarrLinetsky2006}, which is known to be able to capture the empirical evidence of a positive correlation between default probabilities (or credit default swap spreads) and equity volatility.\footnote{See, for example, \cite{Campbell2003}, \cite{ZhangZhouZhu2009}, and \cite{CarrWu2010}.} Moreover, it nests the GBM and CEV models as special cases and, therefore, it also accommodates the aforementioned leverage effect and implied volatility skew stylized facts. The importance of linking equity derivatives markets and credit markets has thus generated a new class of hybrid credit-equity models with the aim of pricing derivatives subject to the risk of default---for other applications of jump to default models, see, for instance, \cite{Nunes2009JFQA}, \cite{ArriagaCarrLinetsky2010}, \cite{LinetskyMendoza-Arriaga2011}, \cite{RuasDiasNunes2013}, \cite{NunesRuasDias2015}, \cite{NunesDiasRuas2017}, and the references contained therein. Moreover, the new algorithms recently provided by \cite{DiasNunes2017} for computing truncated and raw moments of a noncentral $\chi^2$ random variable can be also used on the pricing of barrier options under the JDCEV model considered in \cite{DiasNunesRuas2015}.

The main purpose of this paper is the development of a new analytically tractable NSBF representation for pricing (and hedging) European-style DBKO options which can be applied to the whole class of one-dimensional time-homogeneous diffusions, independently of knowing the corresponding transition density. Similarly to \cite{DavydovLinetsky2003}, we solve the boundary value problem for the parabolic partial differential equation using a classical separation of the variables method. This technique reduces the problem to the determination of the eigenvalues and eigenfunctions of the associated Sturm-Liouville problem. The approach based on the NSBF representation allows one to compute large sets of eigendata with non-deteriorating accuracy. Hence, we are able to calculate the prices for the general time homogeneous diffusion models, not relying on the knowledge of the exact solutions as for example is done in \cite{DavydovLinetsky2003} for the CEV model and \cite{CarrLinetsky2006} and \cite{DiasNunesRuas2015} for the JDCEV model. Therefore, the novel NSBF representation permits the construction of a fast and accurate algorithm for pricing DBKO options and opens its application to other similar problems.\footnote{We recall that \cite{DavydovLinetsky2003} considered also interest rate knock-out options in the \cite{Cox1985b} term structure model. Even though our focus is on equity derivatives, our approach is also suitable for pricing interest rate knock-out options, even in the absence of a closed-form solution for the transition density.}

We note that \cite{CarrLinetsky2006} are able to obtain closed-form solutions for European-style plain-vanilla options, survival probabilities, credit default swap spreads, and corporate bonds in the JDCEV model by exploring the powerful link between CEV and Bessel processes. By adopting the hybrid credit-equity JDCEV architecture modeling framework, \cite{DiasNunesRuas2015} are restricted to the volatility and default intensity specifications that are implicit in the JDCEV model. By contrast, since we do not need to be restricted to such specific modeling assumptions we are able to quickly and accurately price DBKO options for a larger class of models. We illustrate our numerical method on \emph{extended jump to default constant elasticity of volatility} (hereafter, EJDCEV) model, which nests the JDCEV model as a special case.

In summary, our method can be considered as an alternative powerful computational tool for pricing DBKO options. Since we are able to quickly construct the whole value function and not just the price, we can easily observe the behavior of the option price through time and different initial values. Moreover, the NSBF representation also presents an easy way to calculate the derivatives of the value function and consequently the `Greeks' of the option, and thus it can be used for the design of hedging strategies. Given its accuracy, efficiency, and easy implementation, the novel valuation method can be used also for analyzing the empirical performance of models for barrier option valuation under alternative underlying asset pricing dynamics, e.g. \cite{JessenPoulsen2013}.

The remainder of the paper is structured as follows. Section \ref{Sec_Model} sets the general financial framework and defines the boundary value problem. Section \ref{Sec_Solution} provides the main result of the paper (in Proposition \ref{Prop1}): the representation of the solution to the boundary value problem and the price for a DBKO option. Section \ref{Sec_Greeks} illustrates the calculation of the so-called `Greeks'. Next sections are dedicated to the algorithm implementation and numerical examples. First, we present, in section \ref{Sect_Recur}, some recurrence formulas which are more robust and efficient for computation of the coefficients that appear in the direct representations of the value function and its derivatives presented in previous sections. Then, section \ref{Sec_Algorithm} offers the conceptual algorithm for the computation of the price of DBKO options. Section \ref{Sec_Comput} presents numerical experiments for the EJDCEV framework. For illustrative purposes, the analysis is separated into two different horizons: medium (six months) and short (one day). The final section presents the concluding remarks and the possible directions for further research.

\section{Modelling framework\label{Sec_Model}}

This section presents the general financial model for our pricing method and describes the boundary value problem associated to an option contract containing two barrier (knock-out) provisions. We recall that the holder of such a DBKO option has the right to receive, at the expiration date $T$, a payoff $f\left(  y\right)  =\left(  y-K\right)  ^{+}$ in the case of a call option, or $f\left(  y\right)  =\left(  K-y\right)  ^{+}$ for the case of a put option, if the underlying asset price $\left(  Y\right)  $ remains in the range $\left(  L,U\right)  $. The real constants $U>L>0$ are designated by the upper and lower bounds (i.e., the knock-out trigger barrier levels), whereas $K\in\mathbb{R}:\,L<K<U$ is the strike price.\footnote{To simplify the notation, it is assumed that the valuation date of each contract is today (i.e., the current time $t=0$). Moreover, in the case of a knock-out event it is assumed that there is no rebate. Nevertheless, the valuation of rebates can be straightforwardly accomplished using the insights presented in \ref{AppendixA}.}

\subsection{The general financial model setup}

Hereafter, and during the trading interval $\left[0,T\right]$, for some fixed time $T$ ($> 0$), uncertainty is generated by a probability space $\left(\Omega ,\mathcal{G},\mathbb{Q}\right) $, where the equivalent martingale measure $\mathbb{Q}$ associated to the num\'{e}raire \textquotedblleft money market account\textquotedblright\ is taken as given. The price dynamics of the underlying asset is assumed to be governed, under the risk-neutral measure $\mathbb{Q}$, by the time-homogeneous (or time invariant) diffusion
\begin{equation}
dY_{t} = \mu\left(  Y_{t}\right) Y_{t}dt + \sigma\left(  Y_{t}\right)  Y_{y}dB_{t}, \label{NSBF_DBKO_eq1}
\end{equation}
with initial value $Y_{0}=y_{0}$, and where the functions $\mu\left(  y\right) $ and $\sigma\left(  y\right) $ are, respectively, the (state dependent) instantaneous drift and instantaneous volatility (whose regularity properties will be formally defined later in Assumption \ref{Assump_1}), whereas $B_{t} \in \mathbb{R}$ is a standard Wiener process defined under measure $\mathbb{Q}$.

To explain the development of our pricing methodology we consider a European-style DBKO option contract whose payoff at the expiry date $T$ is a function of the single state variable $Y$. Given the contractual clauses of such derivative security, the process is considered on the interval $\left[L,U\right] $, where $L$ and $U$ are, respectively, the lower and upper bounds of the DBKO option contract. The end-points $L$ and $U$ are set to be knock-out boundaries, because if at any time between the initial date of the contract and its expiration either the upper barrier or the lower barrier is hit, then the option contract is canceled (i.e. it is knocked out). We are considering the case with no rebate for the simplicity. However, it is possible to incorporate a rebate value, as shown in the note  in the appendix.

As in \cite{DavydovLinetsky2003}, if any of these end-points is a regular boundary, we adjoin a killing boundary condition at that end-point, sending the process to a cemetery state $\{\Delta\}$ at the first hitting time of the end-point. Consequently, the hitting time for our problem (with two knock-out provisions) is defined by
\begin{equation}
\tau_{\left\{  L,U\right\}  } = \inf\left\{  s\geq t : Y_{s}\notin\left(L,U\right)  \right\}. \notag
\end{equation}

We also consider the possibility that the process may be killed by a sudden jump to $\{\Delta\}$, i.e. the spot price is allowed to jump to zero from the interior of the interval. This implies that the default event forces the option knock-out. There is no recovery value (in the case of a DBKO put) upon default.

This is accomplished by imposing a killing time defined as
\begin{equation}
\tau_{h}=\inf\left\{  s\geq 0:\int_{0}^{s}h\left(  Y_{u}\right) du\geq\mathcal{E}\right\}, \notag
\end{equation}
where $h\left(  y\right)  \geq0$ is the default intensity or the hazard rate, whereas $\mathcal{E}$ is an exponential random variable with unit mean, independent of $Y$ and time. Therefore, the combined stopping time of the process is denoted by
\begin{equation}
\tau = \min \left\{  \tau_{\left\{  L,U\right\}  },\tau_{h}\right\} .
\end{equation}

\begin{remark}
The abuse of notation is in place; to simplify notation, the stopped process set on the domain $\left[ L,U\right]  \cup\left\{  \Delta\right\}  $ is denoted by the same letter $Y$ as the original process. This avoids bloating the paper with many technical details that are standard in the literature---see, for instance, \cite[Section III.3.18]{RogersWilliams1994}, \cite[Section II.4]{BorodinSalminen2002}, \cite[Section 8.2]{Oksendal1995} and \cite{ArriagaCarrLinetsky2010}. We notice that the defaultable asset price process is adapted not to the filtration $\mathbb{F}=\{\mathcal{F}_{t},t\geq 0\}$ generated by the predefault process, but rather to the enlarged filtration $\mathbb{G=}\left\{ \mathcal{G}_{t}:t\geq 0\right\} $, obtained as $\mathcal{G}_{t}=\mathcal{F}_{t}\vee \mathcal{D}_{t}$, where $\left\{ \mathcal{D}_{t},t\geq 0\right\} $ is a default indicator process, with $\mathcal{D}_{t}=1\hspace*{-0.8ex}1_{\{t>\tau_h \}}$.
\end{remark}

As usual, to ensure that the constructed (killed) process remains a martingale it is necessary to set the drift of equation \eqref{NSBF_DBKO_eq1} as
\begin{equation}
\mu\left(  y\right) = \bar{r}\left( y\right)  - \bar{q}\left( y\right)  + h\left( y\right),
\label{eq_drift}
\end{equation}
where $\bar{r}\left( y\right)$ and $\bar{q}\left( y\right)$ are the (time-homogeneous) continuously compounded interest rate and dividend yield.

In summary, the main purpose of this paper is to develop an efficient and flexible pricing methodology for computing risk-neutral expectations of the form
\begin{equation}
v\left(  y,t\right)  =E_{y}\left[  e^{-\int_{t}^{T} \left[\bar{r}\left( Y_{s}\right) + h\left(  Y_{s}\right)\right]  ds} f\left(  Y_{T}\right) \bbOne_{\left\{ \tau>T\right\}  }\right]. \label{NSBF_DBKO_eq2}
\end{equation}
This will be accomplished by applying the NSBF expansion to the associated Sturm-Liouville problem.

\subsection{The boundary value problem}

Let us introduce the differential operator
\begin{equation}
\mathcal{A}=\frac{1}{2}\sigma^{2}\left(
y\right)  y^{2}\partial_{yy} + \mu \left(  y\right)  y\partial_{y} - (\bar{r}\left( y\right) + h\left(  y\right)  ).
\label{NSBF_DBKO_eq3}
\end{equation}

Then, the value function \eqref{NSBF_DBKO_eq2} is the solution of the following boundary value problem
\begin{equation}
\begin{cases}
\mathcal{A}v\left(  y,t\right)=-v_{t}, & \left(  y,t\right) \in \left(  L,U\right) \times \left[ 0,T\right), \\
v\left(  y,T\right)  =f\left(  y\right),  & y\in\left( L,U\right), \\
v\left(  L,t\right)  =v\left(  U,t\right) = 0,   & t \in \left[ 0,T \right].
\end{cases}
\label{eq_BVP}
\end{equation}

The pricing of the DBKO option will be performed by solving problem \eqref{eq_BVP}. For convenience, we rewrite the operator $\mathcal{A}$ in the following form
\[
\mathcal{A}=\frac{1}{w\left(  y\right)  }\left(  \frac{d}{dy}\left(  p\left(
y\right)  \frac{d}{dy}\right)  -q\left(  y\right)  \right)  ,
\]
where
\begin{equation}\label{Defs pwq}
p(  y) = \exp\left(  \int^{y}\frac{2\mu(  s)}{s\sigma^{2}(  s) }ds\right),\quad w(  y) = \frac{2p(  y)  }{\sigma^{2}(y)  y^{2}},\quad\text{and}\quad q(  y) = \left[ \bar{r}( y) + h(  y) \right] w(  y).
\end{equation}

At this point, we can set the needed conditions for the coefficients of the process $Y_{t}$ through the problem \eqref{eq_BVP}. In this illustration paper we are looking only at the regular case, so we will need the following assumptions:
\begin{ass}
\label{Assump_1}The functions $p$, $p^{\prime}$, $q$, $w$ and $w^{\prime}$ are real valued and continuous on $\left[  L,U\right]  $. Additionally, it is assumed that $p^{\prime}$ and $w^{\prime}$ are absolutely continuous and that $p>0$, $\sigma>0$ and $w>0$.
\end{ass}
\begin{ass}
\label{Assump_2}The payoff function $f$ is square integrable.
\end{ass}

Application of Fourier' separation of variables method to the partial differential equation in \eqref{eq_BVP}---see, for example, \cite{Mikhailov1978} and \cite{PinchoverRubinstein2005} for a general exposition of the method and \cite{DavydovLinetsky2003} for financial applications---, leads to the eigenfunction expansion of the value function \eqref{NSBF_DBKO_eq2} as
\begin{equation}
v\left(  y,t\right)  = \sum_{n=1}^{\infty} f_{n}\varphi_{n}\left(  y\right)  e^{-\lambda_{n}(T-t)}, \label{NSBF_DBKO_eq5}
\end{equation}
where the pairs $\left(  \lambda_{n},\varphi_{n}\right)  $ are solutions of the Sturm-Liouville problem
\begin{equation}
\begin{cases}
\left(  p\left(  y\right)  \varphi_{n}^{\prime}\left(  y\right)  \right)
^{\prime}-q\left(  y\right)  \varphi_{n}\left(  y\right)  =-\lambda
_{n}w\left(  y\right)  \varphi_{n}\left(  y\right),  &   y\in\left(
L,U\right) \\
\varphi_{n}\left(  U\right)  =\varphi_{n}\left(  L\right)  =0. &  
\end{cases}
\label{NSBF_DBKO_eq6}
\end{equation}

The functions $\varphi_{n}$ form a complete orthogonal basis for the space
$L_{w}^{2}\left(  \left[  L,U\right]  \right)  $. The coefficients $f_{n}$ are
the Fourier coefficients of the function $f$ relative to the basis $\left\{
\varphi_{n}\right\}  _{n\in\mathbb{N}}$ with scalar product
\begin{equation}
\left\langle g_{1},g_{2}\right\rangle =\int_{L}^{U}g_{1}\left(  s\right)
g_{2}\left(  s\right)  w\left(  s\right)  ds. \label{NSBF_DBKO_eq6_5}%
\end{equation}

The Hilbert space $L_{w}^{2}\left(  \left[  L,U\right]  \right)  $ with the
above defined scalar product is denoted by $H_{w}$. The function $f$ can be explicitly decomposed as
\begin{equation}
f\left(  y\right)  =\sum_{n=1}^{\infty}
f_{n}\varphi_{n}\left(  y\right)%
,
\label{NSBF_DBKO_eq6_7}%
\end{equation}
where $f_{n}$ are the Fourier coefficients of the payoff function $f$ defined as
\[
f_{n}=\frac{\left\langle f,\varphi_{n}\right\rangle }{\left\langle \varphi
_{n},\varphi_{n}\right\rangle }.
\]
We recall that Assumption \ref{Assump_2} guarantees that the series converges to the function $f$ in $L_{w}^{2}$ norm. We note also that Assumption \ref{Assump_1} ensures that problem \eqref{NSBF_DBKO_eq6} is a regular Sturm-Liouville problem. The eigenvalues are real, positive and can be listed as $\lambda_{1}<\lambda_{2}<..<\lambda_{n}<...$, with $\lim_{n\rightarrow\infty}\lambda_{n}=+\infty$.\footnote{Decomposition \eqref{NSBF_DBKO_eq5} and other related topics, such as
properties of the eigenfunctions, can be consulted in \cite[Chapter 10]{BirkhoffRota1989} and \cite[Chapter 7]{StakgoldHolst2011}. The analysis of the spectral decomposition directly applied to finance problems may be found in \cite{Linetsky2004}.}

\begin{remark}
\label{rem_discont}
We further notice that for the put and call barrier options under consideration, problem \eqref{eq_BVP} possesses non-consistent (discontinuous) boundary conditions. For the barrier call option the value function is discontinuous at the point $\left(  U,T\right)  $, i.e.\ $v\left(  U,T\right)  =0\neq\lim_{y\rightarrow U}v\left(  y,T\right) =f\left(  U\right)  =U-K$. A similar observation can be made for the point $\left(  L,T\right)  $ in the barrier put case. Nevertheless, in both cases the value function is still in the space $H_{w}$ and can be represented by its Fourier series \eqref{NSBF_DBKO_eq6_7}.
\end{remark}

\section{An analytical representation through NSBF of the value function\label{Sec_Solution}}

This section presents the pricing formula for the DBKO option using the NSBF representation for the Sturm-Liouville problem \eqref{NSBF_DBKO_eq6} recently proposed by \cite{KravchenkoNavarroTorba2015} for the one-dimensional Schr\"{o}dinger equation---i.e. the case with $w\left(  y\right)  =1$---and then extended to a more general function $w\left(  y\right)  $ in \cite{KravchenkoTorba_Neumann_2016}. In a nutshell, this powerful technique consists in the representation of the solutions of the Sturm-Liouville problem \eqref{NSBF_DBKO_eq6} and their derivatives in terms of NSBF with explicit formulas for the coefficients.

To extend this approach to our option pricing problem, let us first introduce the space $H_{w}^{1,0}$. This is the subspace of the functions $u\in L_{w}^{2}\left(  \left[  L,U\right]  \times\left[0,T\right]  \right)$, that possesses the first derivative $\partial_{x}u$ in the sense of distributions and $\partial_{x}u,u\in L_{w}^{2}\left(  \left[  L,U\right] \times\left[0,T\right]  \right)  $---see, for example, \cite[Chapter III.2]{Mikhailov1978} for further details. Next proposition provides our main theoretical result.

\begin{proposition}
\label{Prop1}Under Assumptions \ref{Assump_1} and \ref{Assump_2}, the value function \eqref{NSBF_DBKO_eq2} is a solution to problem \eqref{eq_BVP} and can be represented as
\begin{equation}
v\left(  y,t\right)  =\sum_{n=1}^{\infty}
f_{n}\left[  \frac{\sin\left(  \omega_{n}l\left(  y\right)  \right)  }%
{\rho\left(  y\right)  }+2\sum_{m=0}^{\infty}
\left(  -1\right)  ^{m}\alpha_{2m+1}\left(  y\right)  j_{2m+1}\left(
\omega_{n}l\left(  y\right)  \right)  \right]  e^{-\omega_{n}^{2}(T-t)}.
\label{NSBF_DBKO_eq7}%
\end{equation}
The series converges in the norm of the space $H_{w}^{1,0}$.\footnote{The price of the DBKO option is given by $v\left(  y_{0},0\right)$ and the corresponding series converges in the norm $L_{w}^{2}\left(  \left[  L,U\right]
\right) $.} Moreover the series converges uniformly with respect to $y\in[L, U]$ and $t\in [0,T_0]\subset [0,T)$.
\end{proposition}

Before providing a formal proof of Proposition \ref{Prop1} and for the sake of completeness, let us first highlight some important details aiming to offer a better exposition. Consider the following identities:
\begin{itemize}
\item The eigenfunctions, solutions to the Sturm-Liouville problem \eqref{NSBF_DBKO_eq6} are\footnote{Note that these functions are not normalized.}
\begin{equation}
\varphi_{n}\left(  y\right)  =\frac{\sin\left(  \omega_{n}l\left(  y\right)
\right)  }{\rho\left(  y\right)  }+2\sum_{m=0}^{\infty}
\left(  -1\right)  ^{m}\alpha_{2m+1}\left(  y\right)  j_{2m+1}\left(
\omega_{n}l\left(  y\right)  \right).
\label{eq_phinSLSol}
\end{equation}

\item The spherical Bessel functions of the first kind, $j_{\nu}\left(  y\right) $, are given by
\[
j_{\nu}\left(  y\right)  =\sqrt{\frac{\pi}{2y}}J_{\nu+\frac{1}{2}}\left(
y\right)  ,
\]
where $J_{\mu}\left(  y\right)  $ are the Bessel functions of the first kind shown in \cite[8.461.1]{GradshteynRyzhik2007}.

\item The function $l\left(  y\right)$ is defined by\footnote{A detailed analysis of the role of this transformation in this decomposition and in the transmutation operators theory can be found in \cite{KravchenkoMorelosTorba2016}.}
\[
l\left(  y\right)  :=\int_{L}^{y}\sqrt{\frac{w\left(  s\right)  }{p\left(
s\right)  }}ds=\sqrt{2}\int_{L}^{y}\frac{1}{s\sigma\left(  s\right)
}ds,\text{ \ \ }y\in\left[  L,U\right].
\]

\item The function $\rho\left(  y\right)$ is defined by
\[
\rho\left(  y\right)  =\left[  p\left(  y\right)  w\left(  y\right)  \right]
^{1/4}=\sqrt{2}\left(  \frac{p\left(  y\right)  }{\sigma\left(  y\right)
y}\right)  ^{1/2},\text{ \ \ \ }y\in\left[  L,U\right].
\]

\item The roots of the eigenvalues $\lambda_{n}$, denoted as $\omega_{n}$, are solutions of the equation
\begin{equation}
\frac{\sin\left(  \omega l\left(  U\right)  \right)  }{\rho\left(  U\right)
}+2\sum_{m=0}^{\infty}
\left(  -1\right)  ^{m}\alpha_{2m+1}\left(  U\right)  j_{2m+1}\left(  \omega
l\left(  U\right)  \right)  =0,\text{ \ \ \ }\omega\in\mathbb{\mathbb{R}}.
\label{NSBF_DBKO_eq7_5}
\end{equation}

\item The functions $\alpha_{n}\left(  y\right)$, $n\ge 0$ are defined as
\begin{equation}
\alpha_{n}\left(  y\right)  =\frac{2n+1}{2}\left(
\sum_{k=0}^{n}
\frac{l_{k,n}\Phi_{k}\left(  y\right)  }{\left(  l\left(  y\right)  \right)
^{k}}-\frac{1}{\rho\left(  y\right)  }\right),  \text{ \ \ \ }y\in\left(
L,U\right].
 \label{NSBF_DBKO_eq7_7}
\end{equation}
The efficient recursive method for computing $\alpha_{n}$ will be presented in Section \ref{Sect_Recur}.

\item $l_{k,n}$ is the coefficient of $x^{k}$ in the Legendre polynomial of order $n$---see, for instance, \cite[Chapter 8]{Abramowitz1972}.

\item $\Phi_{k}\left(  y\right)$ are the formal powers that will be defined in Definition \ref{Def_FormalPowers}.
\end{itemize}

The formal powers $\Phi_{k}\left(  y\right)  $ are constructed on the basis of one non-vanishing solution $g$ of the equation\footnote{For $p$, $p^{\prime}$ and $q$ continuous on $\left[  L,U\right]  $ such solution exists, see \cite[Remark 5]{KravchenkoPorter2010}.}
\begin{equation}
\left(  p\left(  y\right)  g^{\prime}\left(  y\right)  \right)  ^{\prime
}-q\left(  y\right)  g\left(  y\right)  =0,\text{ \ \ \ }y\in\left[
L,U\right]  , \label{NSBF_DBKO_eq8}%
\end{equation}
with an initial condition set as
\begin{equation}
g\left(  L\right)  =\frac{1}{\rho\left(  L\right)  }. \label{NSBF_DBKO_eq9}%
\end{equation}

\begin{defn}
\label{Def_FormalPowers}Let $p$, $q$, $w$ satisfy Assumption \ref{Assump_1} and let $g$ be a non-vanishing solution of equation \eqref{NSBF_DBKO_eq8} that satisfies condition \eqref{NSBF_DBKO_eq9}. Then, the associated formal powers are defined, for $k=0,1,2,...$, as
\[
\Phi_{k}\left(  y\right)  =
\begin{cases}
g\left(  y\right)  Y^{\left(  k\right)  }\left(  y\right),  &   k\text{ odd},\\
g\left(  y\right)  \tilde{Y}^{\left(  k\right)  }\left(  y\right) ,   &
k\text{ even},
\end{cases}
\]
where two families of the auxiliary functions are defined as%
\begin{align*}
Y^{\left(  0\right)  }\left(  y\right)   &  \equiv\tilde{Y}^{\left(  0\right)
}\left(  y\right)  \equiv1,\\
Y^{\left(  k\right)  }\left(  y\right)   &  =
\begin{cases}
k\int_{L}^{y}Y^{\left(  k-1\right)  }\left(  s\right)  \frac{1}{g^{2}\left(
s\right)  p\left(  s\right)  }ds ,   & k\text{ odd}, \\
k\int_{L}^{y}Y^{\left(  k-1\right)  }\left(  s\right)  g^{2}\left(  s\right)
p\left(  s\right)  ds , &   k\text{ even},
\end{cases}
\\
\tilde{Y}^{\left(  k\right)  }\left(  y\right)   &  =
\begin{cases}
k\int_{L}^{y}\tilde{Y}^{\left(  k-1\right)  }\left(  s\right)  g^{2}\left(
s\right)  p\left(  s\right)  ds , &   k\text{ odd}, \\
k\int_{L}^{y}\tilde{Y}^{\left(  k-1\right)  }\left(  s\right)  \frac{1}%
{g^{2}\left(  s\right)  p\left(  s\right)  }ds , &  k\text{ even}.
\end{cases}
\end{align*}
\end{defn}

\begin{remark}
We note that these formal powers arise in the Spectral Parameter Power Series (SPPS) representation for the solution of the Sturm-Liouville problem \eqref{NSBF_DBKO_eq6}. The SPPS method was introduced in \cite{Kravchenko2008}---see also \cite{KravchenkoPorter2010} and \cite{KhmelnytskayaKR2015}.
\end{remark}

Next we provide the formal proof of Proposition \ref{Prop1}.

\begin{pf}[Proposition \ref{Prop1}]
The application of the Fourier separation of the variables method to problem \eqref{eq_BVP} leads to representation \eqref{NSBF_DBKO_eq5}. It is a weak solution of problem \eqref{eq_BVP}---see, for example, \cite[Chapter VI.2, Theorem 3]{Mikhailov1978} and \cite[Chapter 7.1, Theorems 3 and 4]{Evans1998}. Application of \cite[Theorem 3.1]{KravchenkoTorba_Neumann_2016} gives representation
\eqref{NSBF_DBKO_eq7} and guarantees the approximation of the eigenfunctions uniformly in $\omega$. Let us denote by $f_{N}\left(  y\right) = \sum_{n=1}^{N} f_{n}\varphi_{n}\left(  y\right)$ the approximation of the function $f$ of the order $N$. For any $\varepsilon>0$, there is a $N$ such that $\left\Vert f-f_{N}\right\Vert _{L_{w}^{2}}\leq\varepsilon_{N}$,
where $\varepsilon_{N}\rightarrow0$ when $N\rightarrow\infty$. Applying \cite[Chapter VI.2, Theorem 3]{Mikhailov1978}, we have the following estimate
\[
\left\Vert v-v_{N}\right\Vert _{H_{w}^{1,0}}\leq C\left\Vert f-f_{N}%
\right\Vert _{L_{w}^{2}\left[  L,U\right]  }\leq C\varepsilon_{N}. \hspace{2cm}\]

The uniform convergence of the series is due to majorization by decreasing sequence $e^{-\lambda_n T_0}$. \hfill $\blacksquare$
\end{pf}

In summary, Proposition \ref{Prop1} provides a powerful computational technique with the potential to be applied in a wide range of finance applications due to the fact that the NSBF representation can be used as a simple and efficient numerical method. Furthermore, the proposed novel representation is applicable to a large class of option pricing models and it represents not only the price but also the entire value function. This feature allows us to view the behavior of the option price under different initial values for the asset (i.e., to construct the value surface as will be shown in Figure \ref{GraphValueFunction}).

\section{The analytical representation of `Greeks'\label{Sec_Greeks}}

Since Proposition \ref{Prop1} presents an analytical representation of the value function, we are then able to offer a similar representation for its derivatives, commonly known as `Greeks' in the option pricing literature. This should be a useful computational tool for both academics and practitioners, since numerical differentiation is known to be problematic in this kind of problems.

\subsection{Delta}

Let $y_{0}\in \left(L,U\right)$. The Delta can be represented as
\begin{equation}
\Delta=\frac{\partial v}{\partial y}\left(  y_{0},0\right)  =
\sum_{n=1}^{\infty}
f_{n}\varphi_{n}^{\prime}\left(  y_{0}\right)  e^{-\omega_{n}^{2}T},
\label{NSBF_DBKO_eq14}%
\end{equation}
where%
\begin{multline*}
\varphi_{n}^{\prime}\left(  y\right)  =\sqrt{\frac{w\left(  y\right)
}{p\left(  y\right)  }}\left(  \frac{1}{\rho\left(  y\right)  }\left[
G_{2}\left(  y\right)  \sin\left(\omega_{n}l\left(  y\right) \right) +\omega_{n}\cos\left(
\omega_{n}l\left(  y\right)  \right)  \right] + \right.  \\
+\left. 2\sum_{m=0}^{\infty}
\left(  -1\right)  ^{m}\beta_{2m+1}\left(  y\right)  j_{2m+1}\left(
\omega_{n}l\left(  y\right)  \right) \right)- \\
-\frac{\rho^{\prime}\left(  y\right)  }{\rho\left(  y\right)  }\left(
\frac{\sin\left(  \omega_{n}l\left(  y\right)  \right)  }{\rho\left(
y\right)  }+2\sum_{m=0}^{\infty}
\left(  -1\right)  ^{m}\alpha_{2m+1}\left(  y\right)  j_{2m+1}\left(
\omega_{n}l\left(  y\right)  \right)  \right)  ,
\end{multline*}
the functions $G_{2}\left(  y\right)  $ and $\beta_{m}\left(  y\right)  $ are
presented in the next section. The expressions for the $\varphi_{n}^{\prime}$ are adapted from \cite[Section 5]{KravchenkoTorba_Neumann_2016}. The representation \eqref{NSBF_DBKO_eq14} is valid if the function $\frac{\partial v}{\partial y}$ is continuous at $\left(y_{0},0\right)$. The conditions for this can be consulted at \cite[Theorem 12.1]{LadyzhenskayaSU1988}.

\subsection{Vega}

For the calculation of the Vega, we assume that the instantaneous volatility
$\sigma$ is differentiable and $\sigma^{\prime}\left(
y_{0}\right)  \neq0$. Then by the application of the chain rule and the
derivative of the inverse function theorem, we have%
\begin{equation}
\nu=\frac{\partial v}{\partial\sigma}\left(  \sigma\left(  y_{0}\right)
,0\right)  =\frac{\partial v}{\partial y}\left(  y_{0},0\right)  \frac
{1}{\sigma^{\prime}\left(  y_{0}\right)  }=\frac{\Delta}{\sigma^{\prime
}\left(  y_{0}\right)  }. \label{NSBF_DBKO_eq15}%
\end{equation}
For the constant $\sigma$ we cannot apply this formula.\footnote{In particular, see \cite[Section 12.2]{Shaw1998} for the GBM model.}

\subsection{Theta}
The direct differentiation with respect to $t$ of \eqref{NSBF_DBKO_eq7} provides us with a formula for the Theta
\begin{equation}
\theta=\frac{\partial v}{\partial t}\left(  y_{0},0\right)  =\sum_{n=1}^{\infty}
f_{n}\lambda_{n}\varphi_{n}\left(  y_{0}\right)  e^{-\lambda_{n}T}.
\label{NSBF_DBKO_eq16}%
\end{equation}
As in the case of the Delta, it is necessary to assume the continuity of $\frac{\partial v}{\partial t}$ at $\left(y_{0},0 \right)$.
\section{Recurrence formulas for the coefficients $\alpha_{n}\left(  y\right)  $ and $\beta_{n}\left(  y\right) $\label{Sect_Recur}}

For the (efficient and robust) computation of the coefficient functions $\alpha_{n}\left(  y\right) $ and $\beta_{n}\left(  y\right)  $ it is convenient to use recurrence formulas borrowed from \cite{KravchenkoTorba_Neumann_2016}. These formulas increase the robustness of the calculations by solving the numerical issue in \eqref{NSBF_DBKO_eq7_7} related to the fast growth of the Legendre coefficients.

We first introduce
\begin{equation}
A_{n}\left(  y\right)  =l^{n}\left(  y\right)  \alpha_{n}\left(  y\right)
\text{ \ \ \ and \ \ }B_{n}\left(  y\right)  =l^{n}\left(  y\right)  \beta
_{n}\left(  y\right)  . \label{NSBF_DBKO_eq10}%
\end{equation}
The following formulas hold for $n=2,3,...$
\begin{equation}
A_{n}\left(  y\right)  =\frac{2n+1}{2n-3}\left(  l^{2}\left(  y\right)
A_{n-2}\left(  y\right)  +\left(2n-1 \right) g\left(  y\right)  \tilde{\theta}_{n}\left(
y\right)  \right) \notag
\end{equation}
and
\begin{eqnarray}
B_{n}(  y)  & = & \frac{2n+1}{2n-3}\Bigg\{  l^{2}(  y)
B_{n-2}(  y)  +2(2n-1)\Bigg(  \sqrt{\frac{p(  y)
}{w(  y)  }}\left(  g^{\prime}(  y)  \rho(
y)  +g(  y)  \rho^{\prime}(  y)  \right)
\frac{\tilde{\theta}_{n}(  y)  }{\rho(  y)  } \notag
  \\
&& + \frac{\tilde{\eta}_{n}(  y)  }{\rho^{2}(
y)  g(  y)  }\Bigg)  -(2n-1 )  l(
y)  A_{n-2}(  y)  \Bigg\}, \notag
\end{eqnarray}
where
\begin{equation}
\tilde{\theta}_{n}\left(  y\right)  =\int_{A}^{y}\left(  \frac{\tilde{\eta
}_{n}\left(  x\right)  }{\rho^{2}\left(  x\right)  g^{2}\left(  x\right)
}-\frac{l\left(  x\right)  A_{n-2}\left(  x\right)  }{g\left(  x\right)
}\right)  \sqrt{\frac{w\left(  x\right)  }{p\left(  x\right)  }}dx \notag
\end{equation}
and
\begin{equation*}
\tilde{\eta}_{n}(  y) = \int_{A}^{y}\Bigg(  l(  x)
(  g^{\prime}(  x)  \rho(  x)  +g(  x)
\rho^{\prime}(  x)  )  + (  n-1)  \rho(  x)  g(  x)  \sqrt
{\frac{w(  x)  }{p(  x)  }} \Bigg) \rho(  x)
A_{n-2}(  x)  dx.
\end{equation*}
%\begin{eqnarray}
%&&\tilde{\eta}_{n}\left(  y\right) \notag \\
%& = &\int_{A}^{y}\Bigg(  l\left(  x\right)
%\left(  g^{\prime}\left(  x\right)  \rho\left(  x\right)  +g\left(  x\right)
%\rho^{\prime}\left(  x\right)  \right)  + \left(  n-1\right)  \rho\left(  x\right)  g\left(  x\right)  \sqrt
%{\frac{w\left(  x\right)  }{p\left(  x\right)  }} \Bigg) \rho\left(  x\right)
%A_{n-2}\left(  x\right)  dx. \notag
%\end{eqnarray}
The initial values $A_{0}$, $A_{1}$, $B_{0}$ and $B_{1}$ can be calculated from
\begin{align*}
\alpha_{0}\left(  y\right)   &  =\frac{1}{2}\left(  g\left(  y\right)
-\frac{1}{\rho\left(  y\right)  }\right)  \text{ \ \ \ or \ \ \ }A_{0}\left(
y\right)  =\alpha_{0}\left(  y\right), \\
\alpha_{1}\left(  y\right)   &  =\frac{3}{2}\left(  \frac{\Phi_{1}\left(
y\right)  }{l\left(  y\right)  }-\frac{1}{\rho\left(  y\right)  }\right)
\text{ \ \ \ or \ \ \ }A_{1}\left(  y\right)  =\frac{3}{2}\left(  \Phi
_{1}\left(  y\right)  -\frac{1}{\rho\left(  y\right)  l\left(  y\right)
}\right),
\end{align*}
and
\begin{align*}
\beta_{0}\left(  y\right)   &  =\sqrt{\frac{p\left(  y\right)  }{w\left(
y\right)  }}\left(  \alpha_{0}^{\prime}\left(  y\right)  +\frac{\rho^{\prime
}\left(  y\right)  }{\rho\left(  y\right)  }\alpha_{0}\left(  y\right)
\right)  -\frac{G_{1}\left(  y\right)  }{2\rho\left(  y\right)  },\\
\beta_{1}\left(  y\right)   &  =\frac{\alpha_{1}\left(  y\right)  }{l\left(
y\right)  }+\sqrt{\frac{p\left(  y\right)  }{w\left(  y\right)  }}\left(
\alpha_{1}^{\prime}\left(  y\right)  +\frac{\rho^{\prime}\left(  y\right)
}{\rho\left(  y\right)  }\alpha_{1}\left(  y\right)  \right)  -\frac
{3G_{2}\left(  y\right)  }{2\rho\left(  y\right)  },
\end{align*}
with%
\begin{align*}
\alpha_{0}^{\prime}\left(  y\right)   &  =\frac{1}{2}\left(  g^{\prime}(y)
+\frac{\rho^{\prime}\left(  y\right)  }{\rho^{2}\left(  y\right)  }\right), \\
\alpha_{1}^{\prime}\left(  y\right)   &  =\frac{3}{2}\left(  \frac{\left(
g^{\prime}\left(  y\right)  Y^{\left(  1\right)  }\left(  y\right)  +\frac
{1}{g\left(  y\right)  p\left(  y\right)  }\right)  l\left(  y\right)
-g\left(  y\right)  Y^{\left(  1\right)  }\left(  y\right)  \sqrt
{\frac{w\left(  y\right)  }{p\left(  y\right)  }}}{l^{2}\left(  y\right)
}+\frac{\rho^{\prime}\left(  y\right)  }{\rho^{2}\left(  y\right)  }\right)  ,
\end{align*}
and
\begin{equation}
G_{1}\left(  y\right)  =\tilde{h}+G_{2}\left(  y\right),
\label{NSBF_DBKO_eq11}%
\end{equation}
\begin{align}
G_{2}\left((  y\right) & =\frac{1}{2}\int_{L}^{y}\frac{1}{\left(  pw\right)
^{1/4}}\left(  \frac{q}{\left(  pw\right)  ^{1/4}}-\left[  p\left\{  \left(
pw\right)  ^{-1/4}\right\} ' \right]  ^{\prime}\right)  \left(s\right)ds  \nonumber\\
& = \frac{\rho\rho'}{2w}\bigg|_{L}^{y}+\frac{1}{2}\int_{L}^{y}\left[\frac{q}{\rho^{2}}+\frac{\left(\rho'\right)^{2}}{w} \right]\left(s\right)ds,
\label{NSBF_DBKO_eq12}
\end{align}
where
\begin{equation}
\tilde{h}=\sqrt{\frac{\rho\left(  L\right)  }{w\left(  L\right)  }}\left(
\frac{g^{\prime}\left(  L\right)  }{g\left(  L\right)  }+\frac{\rho^{\prime
}\left(  L\right)  }{\rho\left(  L\right)  }\right)  . \label{NSBF_DBKO_eq13}%
\end{equation}

There is a useful practical test for the verification of the coefficients $\alpha_{n}$ and $\beta_{n}$---its details may be consulted in \cite[Equations 7.1-7.3]{KravchenkoTorba_Neumann_2016}---, that is
\begin{align}
\sum_{m=0}^{\infty} \alpha_{m}\left(y\right)&=\frac{\left(G_{1}\left(y\right)+G_{2}\left(y\right)\right) l\left(y\right)}{2\rho\left(y\right)}
\label{eq_ChechNalfa1}\\
\sum_{m=0}^{\infty}\left(-1\right)^{m} \alpha_{m}\left(y\right)&=\frac{\tilde{h}l\left(y\right)}{2\rho\left(y\right)}
\label{eq_ChechNalfa2}
\end{align}
and
\begin{align}
\sum_{m=0}^{\infty} \beta_{m}(y)&=l(y)\left[
\frac{q(y)}{4\rho(y)w(y)}-
\frac{1}{4w(y)} \left[
p(y)\left(\frac{1}{\rho(y)}\right) ^\prime
\right] ^\prime +
\frac{\tilde{h}G_{2}(y)+G_{2}^{2}(y)}{2\rho(y)}
\right],
\label{eq_ChechNbeta1} \\
\sum_{m=0}^{\infty} (-1)^{m}\beta_{m}(y)&=
l(y)\left[\frac{1}{4\rho( y )}
\left( \frac{q(L)}{w(L)} -\frac{\rho(L)}{w(L)}
\left[ p(y)  \left( \frac{1}{\rho(y)} \right)' \right]'\Bigg|_{y=L}
 \right)
+ \frac{\tilde{h}G_{2}\left(y\right)}{2\rho\left(y\right)} \right].
\label{eq_ChechNbeta2}
\end{align}
The relations \eqref{eq_ChechNalfa1} -- \eqref{eq_ChechNbeta2} can also be used as an indicator for the optimal choice of the number $K$ of coefficients in the truncated series \eqref{NSBF_DBKO_eq7} and \eqref{eq_phinSLSol} to include in computations, monitoring the differences between the right sides of equations \eqref{eq_ChechNalfa1} - \eqref{eq_ChechNbeta2} and the partial sums of these.
\begin{remark}
\label{Rem_Alfan}
When computing the coefficients $\alpha_{n}\left(  y\right)  $ and $\beta_{n}\left(  y\right)  $ it is important to properly perform the division by
$l\left(  y\right)  ^{n}$. Here we present a simple scheme that proved to be useful---the detail and the proofs can be consulted in \cite[Section 7]{KravchenkoTorba_Neumann_2016}. Let us first note that the functions $\alpha_{n}$ are crescent functions in the neighborhood of $L$. Then, let $\left\{  y_{i}\right\}_{1\leq i\leq N_{\epsilon}}$ be the ordered set of $N_{\epsilon}$ points of some neighborhood of $L$, $\left[ L,L+\epsilon\right] $, with $y_{1}=L<y_{2}<...<y_{N_{\epsilon}}=L+\epsilon$. For each coefficient function $\alpha_{n}$ consider \footnote{For a function $f:X\rightarrow Y$, the $argmin$ over a subset $S$ of $X$ is defined as $\underset{x\in S\subseteq X}{argmin}\text{ }f\left(  x\right)  :=\left\{  x:x \in S\wedge\forall y\in S:f\left(  y\right)  \geq f\left(  x\right)  \right\}$.}
\begin{equation}
\tilde{y}=\underset{y \in \left\lbrace y_{i}\right\rbrace}{argmin}\text{ }\alpha\left(y\right). \notag
\end{equation}
Let also $k_{0}$ be the index of $\tilde{y}$ (i.e. $y_{k_{0}}=\tilde{y}$). Hence, we can set $\alpha_{n}\left(y\right)=0$ for all $n<k_{0}$. They are larger only due to the numerical error. A similar construction can be performed for the coefficients $\beta_{n}\left(  y\right)$.
\end{remark}

\section{Implementation of the pricing algorithm\label{Sec_Algorithm}}

For the sake of completeness and to better describe important details of our pricing methodology, let us now provide the conceptual steps for implementing our algorithm:

\begin{enumerate}
\item \label{item_first}Compute the coefficients $p$, $q$ and $w$ of the associated Sturm-Liouville problem using \eqref{Defs pwq}.
\item  Create or choose an indefinite integration scheme. In this paper, we have used the Newton-Cotes six point integration rule---see \cite{KravchenkoNavarroTorba2015} for discussions on the use of other possible methods.

\item Construct or find any non-vanishing solution $g$ to equation \eqref{NSBF_DBKO_eq8} that satisfies \eqref{NSBF_DBKO_eq9}. In our implementation, we have used the SPPS method presented in \cite{KravchenkoPorter2010}. For example, if $q\left(  y\right)  \equiv 0$ (as in the case of the standard CEV model), we can choose $g\left(  y\right)  =\frac{1}{\rho\left(L\right)  }$ as a particular solution.

\item\label{item_initG} Construct the formal power $\Phi^{(1)}$ using Definition \ref{Def_FormalPowers}, compute the constant $\tilde{h}$ and the functions $G_{1}\left(y\right)  $ and $G_{2}\left(  y\right)  $ using formulas \eqref{NSBF_DBKO_eq13}, \eqref{NSBF_DBKO_eq11} and \eqref{NSBF_DBKO_eq12}, respectively.

\item \label{item_An}Compute recursively the coefficients $A_{m}\left(  y\right)  $ and $B_{m}\left(  y\right)  $ using the representation highlighted in Section \ref{Sect_Recur}.

\item \label{item_alfan}Compute coefficients $\alpha_{m}\left(  y\right)$ and $\beta_{m}\left(  y\right) $ using equations \eqref{NSBF_DBKO_eq10} and verify them using relations \eqref{eq_ChechNalfa1} -- \eqref{eq_ChechNbeta2} and Remark \ref{Rem_Alfan}. We notice that this procedure can incorporate a test for estimating an optimal $M$ (truncation parameter for the series \eqref{eq_phinSLSol} and for the second sum in the series \eqref{NSBF_DBKO_eq7}) to be used.

\item \label{item_Spectra}Find the eigenvalues $\lambda_n=\omega_n^2$ from equation \eqref{NSBF_DBKO_eq7_5}. Note that the values of the spherical Bessel functions $j_{2m+1}$ for varying indices $m=0,1,\ldots,M$ at the same point $x$ can be computed efficiently using backward-recursive formula, see \cite[Equation 10.1.19]{Abramowitz1972}
\[
j_{m}\left(  x\right)  =\frac{2\left(  m+1\right)  }{x}j_{m+1}\left(
x\right)  -j_{m+2}\left(  x\right).
\]

\item Construct the eigenfunctions of the problem \eqref{NSBF_DBKO_eq6} given by \eqref{eq_phinSLSol}.

\item \label{item_fDecomp} Decompose function $f$ into the Fourier series \eqref{NSBF_DBKO_eq6_7} using the eigenfunctions $\varphi_{n}$.

\item \label{item_vfunction}Construct the function $v$ through a truncated expression \eqref{NSBF_DBKO_eq5}. By $N$ we denote the number of terms in the truncated series.

\item \label{item_Greeks}Calculate the Greeks via expressions \eqref{NSBF_DBKO_eq14} -- \eqref{NSBF_DBKO_eq16}.
\end{enumerate}

\begin{remark}
Notice that the proposed algorithm can be significantly simplified if we are interested only in the price of the option $v\left(  y_{0},0\right)  $. If this is the case, then in steps \ref{item_initG}, \ref{item_An} and \ref{item_alfan} we only need terms relative to $A_{n}$ and $\alpha_{n}\left(  y\right)  $. Moreover, after calculating $f_{n}$ we do not need to keep the eigenfunctions, but only values at the point $y_{0}$. Further, at step \ref{item_vfunction}, we construct only $v\left( y_{0},t\right)  $ and step \ref{item_Greeks} is not necessary.
\end{remark}

\section{\label{Sec_Comput}Computational experiments and particular examples}

In this section we apply the algorithm described in the previous section to the EJDCEV model, whose details will be described next. For illustrative purposes, we have separated the examples in two different time horizons, the medium (six months) and the short (one day). This particular choice will highlight the eigendata needed for the accurate computations and the sensibility of the model to the chosen parameters.

We note that for the regular Sturm-Liouville problems that we are considering in this paper, the asymptotics for the eigenvalue growth is of the order of $n^{2}$, e.g. \cite[Section 2.13]{PolianinZaitsev1970}. In the case of the long horizon the exponential term $e^{-\lambda_{n}\left(  T-t\right)  }$, with $T-t$ of order $\frac{1}{2}$, decays rapidly and the representation \eqref{NSBF_DBKO_eq7} converges quickly. Hence, few eigenvalues and eigenfunctions are needed to secure a good approximation. For the short horizon case, with $T-t$ of order $\frac{1}{360}$, that is analyzed in the second set of numerical experiments, we need more eigenvalues to have an accurate approximation for the option value. We further note that the NSBF method calculates the required eigendata with the same accuracy. This NSBF's important property, of not loosing accuracy for the highest order eigenvalues, makes it an exciting tool for applications to problems requiring large sets of eigenvalues.

Another computational advantage of our method is that there is no need in any two-dimensional grid for computation. The formulas for the steps \ref{item_first}-\ref{item_fDecomp} are one-dimensional.
In order to make the integration errors negligible and to concentrate mainly on the numerical performance we have used an overwhelming number of mesh points (10000) on the interval $[L,U]$ to represent all the functions involved, moreover, even 3000 mesh points produced close results. Once all the coefficients are obtained, the computation of the value function and Delta may be performed only for the arguments $(y,t)\in [L,U]\times [0,T]$ required by application. E.g., option price can be obtained as the value of $v$ at one point $(y_0, 0)$; the value surface requires calculation of the function $v$ on a mesh of about $101\times 101$ points, etc. Even though the main purpose here is to present the ability of the algorithm to be used in a wide variety of modeling contexts and not the optimization speed, the very small computational burden that is required is remarkable.

All the calculations where done in Matlab R2015a.

\subsection{The EJDCEV model}

The volatility specification under the time-homogeneous version of the JDCEV model is given by
\begin{equation}
\sigma\left(  y\right)   =\delta y^{\beta}, \label{eq_JDCEV_sig}
\end{equation}
with $\delta>0$ and $\beta\in\mathbb{R}$. The drift is given by expression \eqref{eq_drift}, with $\bar{r}>0$, $\bar{q}\geq0$ and hazard rate
\[
h\left(  y\right) := h_{1}\left(  y\right)  = b+c\sigma^{2}\left(  y\right)  ,
\]
with $b>0$ and $c>0$. The properties of the constructed diffusion with different parameterizations can be consulted in \cite{CarrLinetsky2006} and \cite{ArriagaCarrLinetsky2010}. The nice feature of the JDCEV model is its analytical tractability, due to the special form of the assumed hazard rate $h_{1}\left(  y\right)  $. The advantage of the NSBF representation is that it allows us to consider different default intensity specifications without any additional effort. Following \cite{Campbell2003}, we have also considered a default intensity specification guaranteeing a positive relationship between the default probability and volatility. Hence, in our variant of the JDCEV model, that we name as the EJDCEV, the default intensity is assumed to be dependent of the constant parameter $\gamma\geq0$ as
\begin{equation}
h\left(  y\right) := h_{1}^{alt}\left(  y\right)  = b+c\sigma^{\gamma}\left(  y\right).
\label{eq_altH}
\end{equation}
It is important to point out that we are not restricted to functions of the form $h_{1}^{alt}$; we can choose any positive continuously differentiable function. This feature allows the possibility of obtaining a wide alternative of default rates when calibrating the model to market prices.

\subsubsection{Medium horizon (6 month)}

For this set-up, we adopt the parameter configuration considered in \cite[Table 2, Panel C]{DiasNunesRuas2015}, that is $y_{0}=100$, $L=90$, $U=120$, $T=0.5$, $\bar{r}=0.1$, $\bar{q}=0$, $b=0.02$, $c=0.5$, and $\sigma_{0}=0.25$. As usual, the scale parameter $\delta$ is calculated, for each $\beta$ value\footnote{Notice that our $\beta$ is equivalent to the $\overline{\beta}$ considered in \cite{DiasNunesRuas2015}.}, through the relation
\begin{equation}
\sigma_{0}=\delta y_{0}^{\beta}, \label{NSBF_DBKO_eq18}
\end{equation}
while keeping the initial instantaneous volatility $\sigma_{0}=0.25$.

\begin{figure}[H]
\centering
  \includegraphics*[trim= 0 0 0 0,scale=1.00]{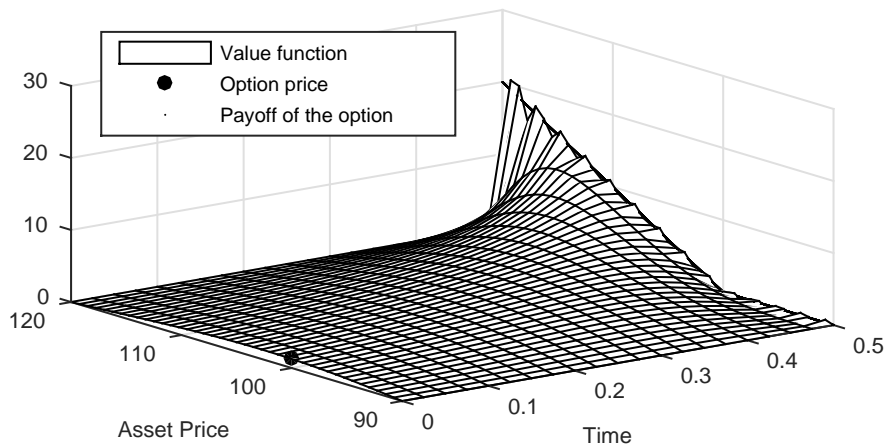}
  \caption{\small{This figure illustrates the value function, the payoff and the price of a European-style DBKO call option, with $y_{0}=100$, $K=100$, $L=90$, $U=120$, $T=0.5$, $\bar{r}=0.1$, $\bar{q}=0$, $b=0.02$, $c=0.5$, $\gamma=2$, $\beta=-1$, and $\sigma_{0}=0.25$.}} \label{GraphValueFunction}
\end{figure}

\begin{figure}[H]
\centering
  \includegraphics*[trim= 0 0 0 0,scale=1.00]{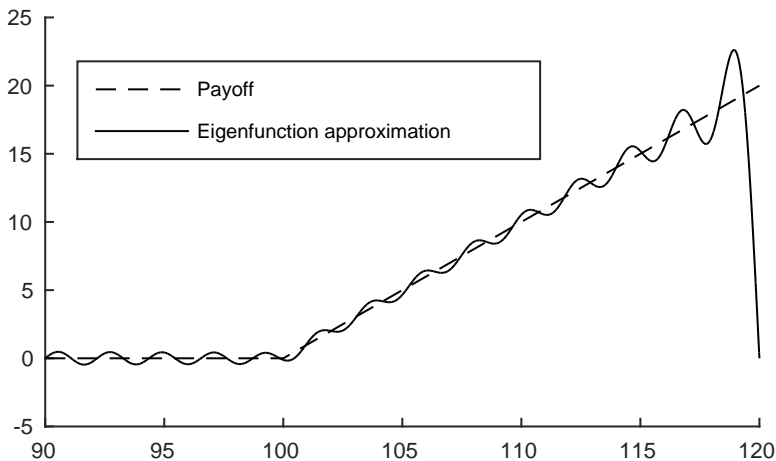}
  \caption{\small{This figure illustrates the payoff function approximation by an eigenfunction expansion for a European-style DBKO call option, with $N=27$ eigenfunctions and $y_{0}=100$, $K=100$, $L=90$, $U=120$, $T=0.5$, $\bar{r}=0.1$, $\bar{q}=0$, $b=0.02$, $c=0.5$, $\gamma=2$, $\beta=-1$, and $\sigma_{0}=0.25$.}} \label{GraffPayoff}
\end{figure}

We notice that the determination of the spectral parameters in step \ref{item_Spectra} was performed by interpolation with a grid of equally distributed 100 points on the interval $\left(0,15\right)$ for the practical purposes and the grid of 1000 points on the interval $\left(0,50\right)$ for the construction of the illustration Table \ref{TableIntradayEigenvalues} and graphs. The practical reasoning is to cut out the eigenvalues $\lambda_{n}>15^2$ due to the term $e^{\lambda_{n}(T-t)}$ in our formulas, this indirectly sets the parameter $N$ somewhere around $6$, as can be observed in Table \ref{TableIntradayEigenvalues}. Figure \ref{GraphValueFunction} illustrates the value function under the JDCEV model using the aforementioned parameters coupled with $K=100$, $\beta=-1$ and $\gamma=2$. Using the same set of parameters, Figure \ref{GraffPayoff} shows the detail of the approximation of the function $f\left(  y\right)=\left(y-K\right)^+ $ at the boundary for $t=T$. It is possible to observe a sharp decline at the boundary $U$, this is the illustration of the Remark \ref{rem_discont}.

%Figure:1 Value Function%%%%%%%%%%%%%%%%%%%%%%%%%%%%%%%%%%%%%%%%%%%%%%%%%%%%%%%%%%%%%%%%%%%%%%%%%%%%%%%%%%%%%%%%%%%%%%%%%%%%%%%%%%%%%%%%%%%%%%%%%%%%%%%%%%%%%%%%%%%%%%

%Figure:2 Payoff Approximation%%%%%%%%%%%%%%%%%%%%%%%%%%%%%%%%%%%%%%%%%%%%%%%%%%%%%%%%%%%%%%%%%%%%%%%%%%%%%%%%%%%%%%%%%%%%%%%%%%%%%%%%%%%%%%%%%%%%%%%%%%%%%%%%%%%%%%%%%%%%%%

Table \ref{Table1} shows the prices of European-style DBKO call and put options and the corresponding Greeks under the EJDCEV model for different moneyness levels with $K \in \{95,100,105\}$, $\beta \in \{0.5, 0.0, -1.0, -2.0\}$ and $\gamma \in \{0,1,2\}$. We further note that the six cases with $\gamma=2$ and $\beta \in \{-1.0, -2.0\}$ originate the values of DBKO put options shown in \cite[Table 2, Panel C]{DiasNunesRuas2015}. A direct comparison reveals that the results are exactly the same (rounded to four decimal places of accuracy), which gives further evidence on the robustness of our algorithm. More importantly, this also allows us to test our methodology under a larger set of volatility and default intensity specifications that until now were not possible to be tackled in the literature.

%Table 1 %%%%%%%%%%%%%%%%%%%%%%%%%%%%%%%%%%%%%%%%%%%%%%%%%%%%%%%%%%%%%%%%%%%%%%%%%%%%%%%%%%%%%%%%%%%%%%%%%%%%%%%%%%%%%%%%%%%%%%%%%%%%%%%%%%%%%%%%%%% %%%%%%%%%%Table 1
%\begin{landscape}
\renewcommand{\LTcapwidth}{6.5in}
%\begin{landscape}
\begin{center}
\renewcommand\thefootnote{\thempfootnote}
\renewcommand\footnoterule{}
\renewcommand{\baselinestretch}{1.1}\small\normalsize
\scalefont{0.75}

\begin{longtable}{rrrrrrrrrrrrrrr}
\caption{Prices and `Greeks' of European-style DBKO options.}\\

\hline
 & & & & \multicolumn{ 5}{c}{{Call Option: $f\left(y\right)=\left(y-K\right)^{+}$}} & & \multicolumn{ 5}{c}{{Put Option: $f\left(y\right)=\left(K-y\right)^{+}$}}\\
 \cline{5-9} \cline{11-15}

\multicolumn{ 3}{c}{Parameters} & & \multicolumn{ 1}{c}{Price} & &  \multicolumn{3}{c}{Greeks} & & \multicolumn{ 1}{c}{Price} & &  \multicolumn{3}{c}{Greeks} \\ \cline{1-3} \cline{5-5} \cline{7-9} \cline{11-11} \cline{13-15}
$K$ &  $\beta$ &  $\gamma$ & & $v\left(100,0\right)$ &  & \multicolumn{ 1}{c}{$\Delta$} & $\nu$ & $\theta$&    & $v\left(100,0\right)$ &            &   $\Delta$ & $\nu$ & $\theta$\\

\endfirsthead
 \caption[]{(continued)}\\
\hline
\multicolumn{ 3}{c}{Parameters} & & \multicolumn{ 1}{c}{Price} & &  \multicolumn{3}{c}{Greeks} & & \multicolumn{ 1}{c}{Price} & &  \multicolumn{3}{c}{Greeks} \\ \cline{1-3} \cline{5-5} \cline{7-9} \cline{11-11} \cline{13-15}
$K$ &  $\beta$ &  $\gamma$ & & $v\left(100,0\right)$ &  & \multicolumn{ 1}{c}{$\Delta$} & $\nu$ & $\theta$&    & $v\left(100,0\right)$ &            &   $\Delta$ & $\nu$ & $\theta$\\

\hline
 \endhead
%Footers
\multicolumn{15}{r}{\footnotesize \textsl{\tiny{continues on next page}}}\\
 \endfoot
 \endlastfoot

\hline
95  & 0.5  & 0 &  & 0.7314 &  & -0.0332 & -26.5669 & 4.9860 &  & 0.0029 &  & -0.0001 & -0.1101 & 0.0201 \\
95  & 0.5  & 1 &  & 1.5057 &  & 0.0179  & 14.3442  & 6.5544 &  & 0.0168 &  & 0.0002  & 0.1364  & 0.0744 \\
95  & 0.5  & 2 &  & 1.5572 &  & 0.0417  & 33.3312  & 6.3003 &  & 0.0222 &  & 0.0006  & 0.4438  & 0.0912 \\
95  & 0.0  & 0 &  & 0.7163 &  & -0.0319 &          & 5.0712 &  & 0.0023 &  & -0.0001 &         & 0.0166 \\
95  & 0.0  & 1 &  & 1.6417 &  & 0.0251  &          & 7.0686 &  & 0.0148 &  & 0.0002  &         & 0.0652 \\
95  & 0.0  & 2 &  & 1.7117 &  & 0.0518  &          & 6.7849 &  & 0.0198 &  & 0.0006  &         & 0.0802 \\
95  & -1.0 & 0 &  & 0.6905 &  & -0.0300 & 12.0097  & 5.2996 &  & 0.0014 &  & -0.0001 & 0.0263  & 0.0111 \\
95  & -1.0 & 1 &  & 1.9733 &  & 0.0432  & -17.2851 & 8.2401 &  & 0.0114 &  & 0.0002  & -0.0865 & 0.0496 \\
95  & -1.0 & 2 &  & 2.0860 &  & 0.0771  & -30.8585 & 7.8538 &  & 0.0157 &  & 0.0005  & -0.2135 & 0.0615 \\
95  & -2.0 & 0 &  & 0.6421 &  & -0.0280 & 5.5973   & 5.3842 &  & 0.0008 &  & 0.0000  & 0.0078  & 0.0071 \\
95  & -2.0 & 1 &  & 2.3959 &  & 0.0675  & -13.5059 & 9.5993 &  & 0.0087 &  & 0.0002  & -0.0419 & 0.0375 \\
95  & -2.0 & 2 &  & 2.5570 &  & 0.1107  & -22.1395 & 9.0265 &  & 0.0123 &  & 0.0005  & -0.0964 & 0.0469 \\
\hline
100 & 0.5  & 0 &  & 0.4568 &  & -0.0207 & -16.5434 & 3.1114 &  & 0.0270 &  & -0.0013 & -1.0175 & 0.1860 \\
100 & 0.5  & 1 &  & 0.8695 &  & 0.0105  & 8.4109   & 3.7784 &  & 0.1307 &  & 0.0014  & 1.0802  & 0.5772 \\
100 & 0.5  & 2 &  & 0.8778 &  & 0.0237  & 18.9282  & 3.5444 &  & 0.1655 &  & 0.0042  & 3.3387  & 0.6801 \\
100 & 0.0  & 0 &  & 0.4561 &  & -0.0202 &          & 3.2256 &  & 0.0218 &  & -0.0010 &         & 0.1563 \\
100 & 0.0  & 1 &  & 0.9700 &  & 0.0150  &          & 4.1676 &  & 0.1181 &  & 0.0016  &         & 0.5189 \\
100 & 0.0  & 2 &  & 0.9881 &  & 0.0301  &          & 3.9064 &  & 0.1517 &  & 0.0043  &         & 0.6143 \\
100 & -1.0 & 0 &  & 0.4571 &  & -0.0198 & 7.9187   & 3.5041 &  & 0.0137 &  & -0.0006 & 0.2535  & 0.1071 \\
100 & -1.0 & 1 &  & 1.2159 &  & 0.0269  & -10.7784 & 5.0594 &  & 0.0962 &  & 0.0018  & -0.7382 & 0.4167 \\
100 & -1.0 & 2 &  & 1.2574 &  & 0.0469  & -18.7457 & 4.7137 &  & 0.1272 &  & 0.0044  & -1.7458 & 0.4979 \\
100 & -2.0 & 0 &  & 0.4385 &  & -0.0190 & 3.8045   & 3.6716 &  & 0.0082 &  & -0.0004 & 0.0781  & 0.0709 \\
100 & -2.0 & 1 &  & 1.5313 &  & 0.0437  & -8.7342  & 6.1011 &  & 0.0779 &  & 0.0019  & -0.3774 & 0.3328 \\
100 & -2.0 & 2 &  & 1.6006 &  & 0.0699  & -13.9770 & 5.6109 &  & 0.1059 &  & 0.0042  & -0.8350 & 0.4012 \\
\hline
105 & 0.5  & 0 &  & 0.2314 &  & -0.0104 & -8.3529  & 1.5750 &  & 0.1004 &  & -0.0047 & -3.7580 & 0.6899 \\
105 & 0.5  & 1 &  & 0.4019 &  & 0.0049  & 3.9499   & 1.7435 &  & 0.4133 &  & 0.0044  & 3.4963  & 1.8212 \\
105 & 0.5  & 2 &  & 0.3948 &  & 0.0107  & 8.5777   & 1.5909 &  & 0.5054 &  & 0.0129  & 10.2860 & 2.0713 \\
105 & 0.0  & 0 &  & 0.2373 &  & -0.0105 &          & 1.6764 &  & 0.0828 &  & -0.0038 &         & 0.5924 \\
105 & 0.0  & 1 &  & 0.4611 &  & 0.0073  &          & 1.9763 &  & 0.3842 &  & 0.0053  &         & 1.6823 \\
105 & 0.0  & 2 &  & 0.4570 &  & 0.0140  &          & 1.8016 &  & 0.4762 &  & 0.0137  &         & 1.9220 \\
105 & -1.0 & 0 &  & 0.2522 &  & -0.0109 & 4.3457   & 1.9300 &  & 0.0544 &  & -0.0025 & 0.9987  & 0.4244 \\
105 & -1.0 & 1 &  & 0.6092 &  & 0.0137  & -5.4756  & 2.5241 &  & 0.3317 &  & 0.0065  & -2.5939 & 1.4293 \\
105 & -1.0 & 2 &  & 0.6129 &  & 0.0230  & -9.2182  & 2.2859 &  & 0.4227 &  & 0.0147  & -5.8633 & 1.6465 \\
105 & -2.0 & 0 &  & 0.2536 &  & -0.0109 & 2.1851   & 2.1184 &  & 0.0342 &  & -0.0016 & 0.3217  & 0.2940 \\
105 & -2.0 & 1 &  & 0.8049 &  & 0.0233  & -4.6594  & 3.1840 &  & 0.2853 &  & 0.0070  & -1.4099 & 1.2092 \\
105 & -2.0 & 2 &  & 0.8184 &  & 0.0361  & -7.2223  & 2.8436 &  & 0.3736 &  & 0.0149  & -2.9815 & 1.4039 \\

\hline

\label{Table1}
\end{longtable}
\end{center}
\vspace{-1.5cm}
\footnotesize{This table shows the prices of European-style DBKO call and put options and the corresponding Greeks under the EJDCEV model, with $y_{0}=100$, $K \in \{95,100,105\}$, $L=90$, $U=120$, $T=0.5$, $\bar{r}=0.1$, $\bar{q}=0$, $b=0.02$, $c=0.5$, $\gamma \in \{0,1,2\}$, $\beta \in \{0.5, 0.0, -1.0, -2.0\}$, and $\sigma_{0}=0.25$.}

\normalsize
\vspace{.5cm}

\begin{figure}[h]
  \centering
    \begin{tabular}{cc}
      \resizebox{3in}{!}{\includegraphics{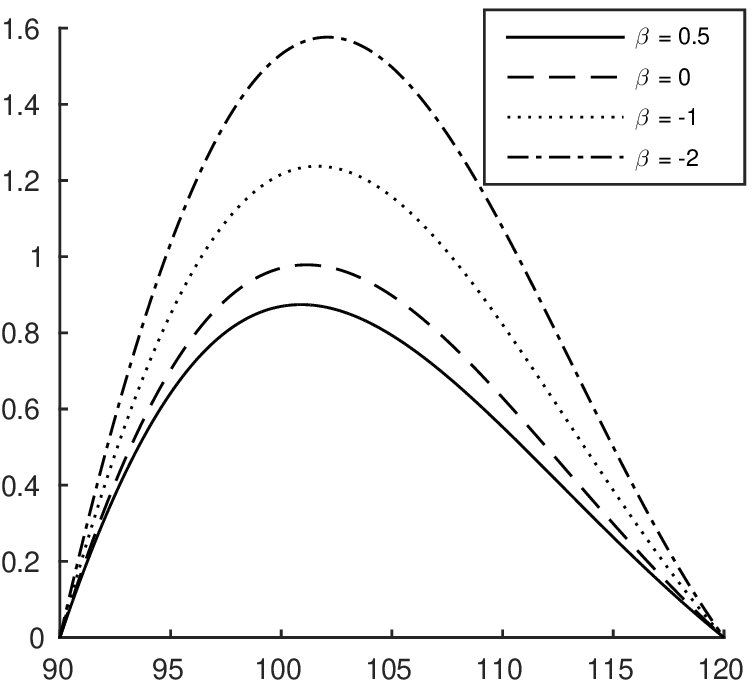}} &
      \resizebox{3in}{!}{\includegraphics{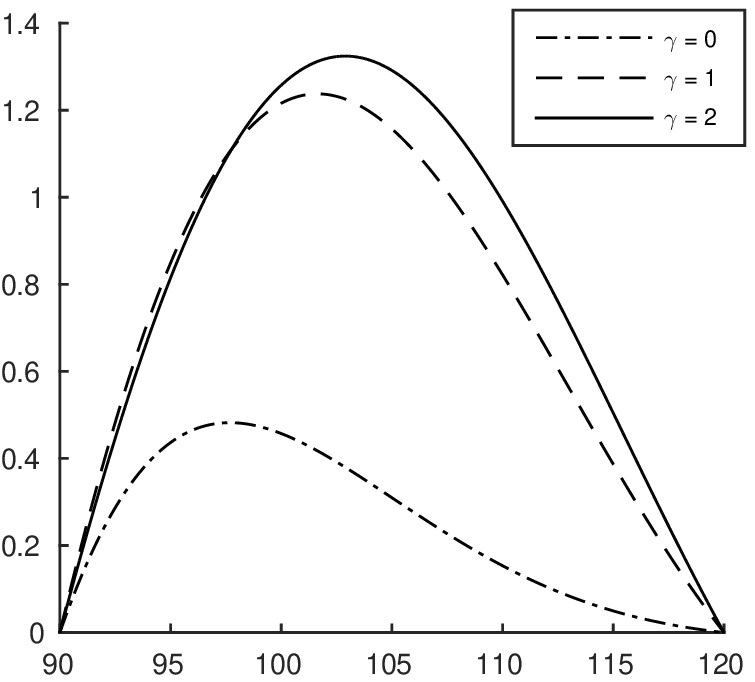}} \\
    \end{tabular}
    \caption{\small{This figure prices European-style DBKO call options for different initial asset values $y_0$. The left-hand side plot sets $\gamma=1$ for different $\beta$ values. The right-hand side plot sets $\beta=-1$ for different $\gamma$ values. The remaining parameters are: $K=100$, $L=90$, $U=120$, $T=0.5$, $\bar{r}=0.1$, $\bar{q}=0$, $b=0.02$, $c=0.5$, and $\sigma_{0}=0.25$.}}
    \label{Fig_DiffBetaGamma}
\end{figure}

Figure \ref{Fig_DiffBetaGamma} shows prices of European-style DBKO call options for different initial asset values $S_0$. The left-hand side plot sets $\gamma=1$ for different $\beta$ values. The right-hand side plot sets $\beta=-1$ for different $\gamma$ values. We note that with the chosen parametrization for this model, the term $e^{\lambda_{n}T}$ decays very rapidly and thus we only need to compute few eigenvalues to obtain accurate prices.

%Figure 3 (Two Graphs side by side)
% Needs
%\usepackage{caption}

%% End Figure 3

\subsubsection{Short horizon (1 day)}

In the case of the short horizon, the time is of the order $\frac{1}{360}$ and, hence, the term $e^{-\lambda_{n}\left(  T-t\right)  }$ decays much slower as $n$ grows. For this case, in step \ref{item_Spectra}, we have used 1000 points for the interval $\left(0,100\right)$. Some values are presented in Table \ref{TableIntradayEigenvalues}. In order to illustrate the convergence and the necessity of calculating accurately a
significant number of eigenvalues, we introduce the contribution of the partial sum from equation \eqref{NSBF_DBKO_eq5}, defined as
\begin{equation}
Contrib\left(  n_{1},n_{2}\right)  =\sum_{n=n_{1}}^{n_{2}}
f_{n}\varphi_{n}e^{-\lambda_{n}(T-t)}.
\label{eq_Contribution}
\end{equation}
We observe, in Table \ref{TableIntradayPrice}, that the value of the
parameter $\beta$ under the short horizon does not have much influence on the
price. However, the $\gamma$ parameter associated with default intensity is relevant.
It is important to note that although the prices of the options for different
$\beta$ values differ slightly, the corresponding Sturm-Liouville problems are very
different. This can be observed in Tables \ref{TableIntradayEigenvalues}
and \ref{TableIntradayContribution}. The observation of the Table \ref{TableIntradayContribution} induces the choice of $N$ around $45$.

%%%%%%%%Table 2: Eigenvalues %%%%%%%%%%%%%%%%%%%%%%%%%%%%%%%%%%%%%%%%%%%%%%%%%%%%%%%%%%%%%%%%%%%%%%%%%%%%%%%%%%%%%%%%%%%%%%%%%%%%%%%%%%%%%%%%%%%%%%%%%% %%%%%%%%%%
\renewcommand{\LTcapwidth}{6.5in}
%\begin{landscape}
\begin{center}
%\begin{minipage}{1.00\linewidth}
\renewcommand\thefootnote{\thempfootnote}
\renewcommand\footnoterule{}
\renewcommand{\baselinestretch}{1.1}\small\normalsize
\scalefont{0.75}

\begin{longtable}{rrrrrr}
\caption{Eigenvalues.}\\
%Definitions for long tables
%Headers of long tables
\hline
&         & \multicolumn{ 4}{c}{Parameters} \\
 \cline{2-6}
$n$ & & $\beta=1,\gamma=1$ & $\beta=1,\gamma=2$ &  $\beta=-2,\gamma=1$ & $\beta=-2,\gamma=2$ \\

\hline

1  &  & 4.4047    & 4.1314    & 4.0997    & 3.6155    \\
6  &  & 144.3068  & 144.0338  & 112.8959  & 112.4098  \\
11 &  & 484.0679  & 483.7949  & 377.105   & 376.6189  \\
16 &  & 1023.6885 & 1023.4155 & 796.731   & 796.2449  \\
21 &  & 1763.1687 & 1762.8956 & 1371.7741 & 1371.288  \\
26 &  & 2702.5083 & 2702.2352 & 2102.2343 & 2101.7481 \\
31 &  & 3841.7073 & 3841.4343 & 2988.1115 & 2987.6253 \\
36 &  & 5180.7659 & 5180.4929 & 4029.4057 & 4028.9196 \\
41 &  & 6719.684  & 6719.411  & 5226.117  & 5225.6309 \\

\hline

\label{TableIntradayEigenvalues}
\end{longtable}
\end{center}
\vspace{-1.5cm}
\footnotesize{This table shows the eigenvalues for different $\gamma$ and $\beta$ parameters, with $y_{0}=100$, $K=100$, $L=90$, $U=120$, $T=0.5$, $\bar{r}=0.1$, $\bar{q}=0$, $b=0.02$, $c=0.5$, and $\sigma_{0}=0.25$.}

%% End Table 2

%\newpage
% Table 3: Call Option Intraday %%%%%%%%%%%%%%%%%%%%%%%%%%%%%%%%%%%%%%%%%
\renewcommand{\LTcapwidth}{6.5in}
%\begin{landscape}
\begin{center}
%\begin{minipage}{1.00\linewidth}
\renewcommand\thefootnote{\thempfootnote}
\renewcommand\footnoterule{}
\renewcommand{\baselinestretch}{1.1}\small\normalsize
\scalefont{0.75}

\begin{longtable}{lrrrr}
\caption{Prices for one-day to maturity European-style DBKO calls.}\\
%Definitions for long tables
%Headers of long tables
\hline
 &$\gamma=3$ & $\gamma=2$ &  $\gamma=1$ & $\gamma=0$ \\

\hline

$\beta=-2$  & 0.54297 & 0.54622 & 0.55950 & 0.61518 \\
$\beta=1$ & 0.54300 & 0.54634 & 0.55976 & 0.61516 \\
\hline

\label{TableIntradayPrice}
\end{longtable}
\end{center}
\vspace{-1.5cm}
\footnotesize{This table shows prices for one-day to maturity European-style DBKO call options for different $\gamma$ and $\beta$ parameters, with $y_{0}=100$, $K=100$, $L=90$, $U=120$, $T=1/360$, $\bar{r}=0.1$, $\bar{q}=0$, $b=0.02$, $c=0.5$, and $\sigma_{0}=0.25$.}

% End Table 3

%Table 4: Contributions %%%%%%%%%%%%%%%%%%%%%%%%%%%%%%%%%%%%%%%%%%%%%%%%%%%%%%%%%%%%%%%%%%%%%%%%%%%%%%%%%%%%%%%%%%%%%%%%%%%%%%%%%%%%%%%%%%%%%%%%%% %%%%%%%%%%
\renewcommand{\LTcapwidth}{6.5in}
%\begin{landscape}
\begin{center}
%\begin{minipage}{1.00\linewidth}
\renewcommand\thefootnote{\thempfootnote}
\renewcommand\footnoterule{}
\renewcommand{\baselinestretch}{1.1}\small\normalsize
\scalefont{0.75}

\begin{longtable}{cccccc}
\caption{Contribution values for one-day to maturity European-style DBKO calls.}\\
%Definitions for long tables
%Headers of long tables
\hline
$n$ & & $\beta=-2,\gamma=2$ &  $\beta=-2,\gamma=1$ & $\beta=1,\gamma=2$ & $\beta=1,\gamma=1$\\

\endfirsthead
 \caption[]{(continued)}\\
\hline
$n$ & & $\beta=-2,\gamma=2$ &  $\beta=-2,\gamma=1$ & $\beta=1,\gamma=2$ & $\beta=1,\gamma=1$\\

\hline
 \endhead
%Footers
\multicolumn{6}{r}{\footnotesize \textsl{\tiny{continues on next page}}}\\
 \endfoot
 \endlastfoot

\hline

1-5 &           & -0.94494 & -1.60020 & 1.54180  & 1.77004  \\
6-10&           & 1.81670  & 2.60534  & -1.15441 & -1.41771 \\
10-15&          & -0.23014 & -0.31208 & 0.19023  & 0.24668  \\
16-20&          & -0.10622 & -0.14909 & -0.03420 & -0.04298 \\
21-25 &         & 0.00934  & 0.01343  & 0.00311  & 0.00400  \\
26-30 &         & 0.00157  & 0.00224  & -0.00021 & -0.00026 \\
31-35 &         & -0.00010 & -0.00014 & 0.00001  & 0.00001  \\
36-40 &         & -0.00001 & -0.00001 & 0.00000  & 0.00000  \\
41-45 &         & 0.00000  & 0.00000  & 0.00000  & 0.00000  \\
\textgreater45 & & 0.00000  & 0.00000  & 0.00000  & 0.00000  \\
\cline{1-6}
Price    &      & 0.54622  & 0.55950  & 0.54634  & 0.55976\\

\hline

\label{TableIntradayContribution}
\end{longtable}
\end{center}
\vspace{-1.5cm}
\footnotesize{This table shows the value of the contribution defined in equation \eqref{eq_Contribution} for one-day to maturity European-style DBKO call options for different $\gamma$ and $\beta$ parameters, with $y_{0}=100$, $K=100$, $L=90$, $U=120$, $T=1/360$, $\bar{r}=0.1$, $\bar{q}=0$, $b=0.02$, $c=0.5$, and $\sigma_{0}=0.25$.}

% End Table 4
\normalsize

\section{Concluding remarks and future research\label{SectionConclusions}}

This paper provides a new methodology for pricing (and hedging) European-style DBKO options via the application of the NSBF decomposition of the Sturm-Liouville equation associated to the corresponding boundary value problem. The illustration of the method was done through the EJDCEV model. The modeling techniques applied in this paper open several avenues for future research. For instance, it should be possible to apply the NSBF decomposition and similar constructions to other singular problems that naturally appear in many financial applications, e.g. plain-vanilla options (unbounded domains), default cases (singularities in the coefficients) and others. It would also be interesting to apply the method to calibrate a parametric curve of parameters to real market data. Finally, it has also the potential to be applied to stopping time problems and related subjects.

\section*{Acknowledgements}

Research was supported by CONACYT, Mexico via the project 222478. The financial support provided by the Funda\c{c}\~{a}o para a Ci\^{e}ncia e Tecnologia (Grant UID/GES/00315/2013) is also gratefully acknowledged. The first named author would like to express his gratitude to the excellence scholarship granted by the Mexican Government via the Ministry of Foreign Affairs which gave him the opportunity to develop this work during his stay in the CINVESTAV, Mexico.

\appendix

%%%%%%%%%%%%%%%%%%%%%%%%%%%%%%%%%%%%%%%%%%%%%%%%%%%%%%%%%%%%%%%%%%%%%%%%%%
% Beginning of appendix

% ------------------------------------------------------------------------
% Definitions for the appendix
% ------------------------------------------------------------------------

% Beginning of appendix A

\renewcommand{\theequation}{\Alph{section}.\arabic{equation}}
\setcounter{equation}{0}
\renewcommand{\thesection}{\appendixname\ \Alph{section}}

\section{Pricing rebates\label{AppendixA}}

Consider the case of the call option with rebate $R>0$. The upper boundary
condition for the problem \eqref{eq_BVP} changes to
\[
v\left(  U,t\right)  =R.
\]
The boundary conditions \eqref{eq_BVP} become non-homogeneous. For the
direct application of the presented method we have to first transform our
value function. Let us define the new function
\[
\tilde{v}\left(  y,t\right)  =v\left(  y,t\right)  -\frac{y-L}{U-L}R,
\]
which satisfies the following boundary problem%
\begin{equation}
\left\{
\begin{array}
[c]{cll}%
\left(  i\right)  & \mathcal{A}\tilde{v}+\mathcal{A}\left(  \frac{y-L}%
{U-L}R\right)  =-\tilde{v}_{t}, & y\times t\in\left(  L,U\right)
\times\left(  0,T\right) \\
\left(  ii\right)  & \tilde{v}\left(  L,t\right)  =\tilde{v}\left(
U,t\right)  =0, & t\in\left[  0,T\right] \\
\left(  iii\right)  & \tilde{v}\left(  y,T\right)  =d\left(  y\right)  , &
y\in\left(  L,U\right)
\end{array}
\right.,
\end{equation}
with homogeneous boundary conditions. The details can be consulted in
\cite[Ch. 6.6]{PinchoverRubinstein2005}. The interesting observation is that
from a mathematical point of view the solution $v\left(  y,t\right)  $ becomes
smoother if $R=U-K$, i.e. the boundary conditions become consistent.

%\section*{References}

%\bibliographystyle{model2-names}
%\bibliographystyle{abbrv}
%\bibliography{jcdbib070817}
%\bibliography{BibTest1bib}
%\bibliography{BibCombined_v2}

\begin{thebibliography}{10}

\bibitem{Abramowitz1972}
M.~Abramowitz and I.~A. Stegun.
\newblock {\em Handbook of Mathematical Functions}.
\newblock Dover, New York, 1972.

\bibitem{BirkhoffRota1989}
G.~Birkhoff and G.-C. Rota.
\newblock {\em Ordinary Differential Equations}.
\newblock John Wiley \& Sons, US, 4th edition, 1989.

\bibitem{BorodinSalminen2002}
A.~N. Borodin and P.~Salminen.
\newblock {\em Handbook of Brownian Motion -- Facts and Formulae}.
\newblock Birkhauser, Basel, 2nd edition, 2002.

\bibitem{BoyleTian1999}
P.~P. Boyle and Y.~Tian.
\newblock Pricing lookback and barrier options under the {CEV} process.
\newblock {\em Journal of Financial and Quantitative Analysis}, 34(2):241--264,
  1999.

\bibitem{BuchenKonstandatos2009}
P.~Buchen and O.~Konstandatos.
\newblock A new approach to pricing double-barrier options with arbitrary
  payoffs and exponential boundaries.
\newblock {\em Applied Mathematical Finance}, 16(6):497--515, 2009.

\bibitem{Campbell2003}
J.~Y. Campbell and G.~B. Taksler.
\newblock Equity volatility and corporate bond yields.
\newblock {\em Journal of Finance}, 58(6):2321--2349, 2003.

\bibitem{CarrLinetsky2006}
P.~Carr and V.~Linetsky.
\newblock A jump to default extended {CEV} model: An application of {B}essel
  processes.
\newblock {\em Finance and Stochastics}, 10(3):303--330, 2006.

\bibitem{CarrWu2010}
P.~Carr and L.~Wu.
\newblock Stock options and credit default swaps: A joint framework for
  valuation and estimation.
\newblock {\em Journal of Financial Econometrics}, 8(4):409--449, 2010.

\bibitem{Cox1975}
J.~C. Cox.
\newblock Notes on option pricing {I}: Constant elasticity of variance
  diffusions.
\newblock 1975.
\newblock Working Paper, Stanford University. Reprinted in {\it Journal of
  Portfolio Management}, 23 (1996), 15-17.

\bibitem{Cox1985b}
J.~C. Cox, J.~E. Ingersoll, Jr., and S.~A. Ross.
\newblock A theory of the term structure of interest rates.
\newblock {\em Econometrica}, 53(2):385--408, 1985.

\bibitem{DavydovLinetsky2001}
D.~Davydov and V.~Linetsky.
\newblock Pricing and hedging path-dependent options under the {CEV} process.
\newblock {\em Management Science}, 47(7):949--965, 2001.

\bibitem{DavydovLinetsky2003}
D.~Davydov and V.~Linetsky.
\newblock Pricing options on scalar diffusions: An eigenfunction expansion
  approach.
\newblock {\em Operations Research}, 51(2):185--209, 2003.

\bibitem{DiasNunes2017}
J.~C. Dias and J.~P. Nunes.
\newblock Universal recurrence algorithm for computing {N}uttall, generalized
  {M}arcum and incomplete {T}oronto functions and moments of a noncentral
  $\chi^{2}$ random variable.
\newblock {\em European Journal of Operational Research.}, 2017.
\newblock Forthcoming.

\bibitem{DiasNunesRuas2015}
J.~C. Dias, J.~P. Nunes, and J.~P. Ruas.
\newblock Pricing and static hedging of {E}uropean-style double barrier options
  under the jump to default extended {CEV} model.
\newblock {\em Quantitative Finance}, 15(12):1995--2010, 2015.

\bibitem{Evans1998}
L.~C. Evans.
\newblock {\em Partial Differential Equations}.
\newblock Graduate Studies in Mathematics, Volume 19. American Mathematical
  Society, 1998.

\bibitem{GemanYor1996}
H.~Geman and M.~Yor.
\newblock Pricing and hedging double-barrier options: A probabilistic approach.
\newblock {\em Mathematical Finance}, 6(4):365--378, 1996.

\bibitem{GradshteynRyzhik2007}
I.~S. Gradshteyn and I.~M. Ryzhik.
\newblock {\em Table of Integrals, Series, and Products}.
\newblock Academic Press, Burlington, MA, 7th edition, 2007.

\bibitem{JessenPoulsen2013}
C.~Jessen and R.~Poulsen.
\newblock Empirical performance of models for barrier option valuation.
\newblock {\em Quantitative Finance}, 13(1):1--11, 2013.

\bibitem{KhmelnytskayaKR2015}
K.~V. Khmelnytskaya, V.~V. Kravchenko, and H.~C. Rosu.
\newblock Eigenvalue problems, spectral parameter power series, and modern
  applications.
\newblock {\em Mathematical Methods in the Applied Sciences},
  38(10):1945--1969, 2015.

\bibitem{Kravchenko2008}
V.~V. Kravchenko.
\newblock A representation for solutions of the {S}turm-{L}iouville equation.
\newblock {\em Complex Variables and Elliptic Equations}, 53(8):775--789, 2008.

\bibitem{KravchenkoMorelosTorba2016}
V.~V. Kravchenko, S.~Morelos, and S.~M. Torba.
\newblock Liouville transformation, analytic approximation of transmutation
  operators and solution of spectral problems.
\newblock {\em Appl.\ Math.\ Comp.}, 273:321--336, 2016.

\bibitem{KravchenkoNavarroTorba2015}
V.~V. Kravchenko, L.~J. Navarro, and S.~M. Torba.
\newblock Representation of solutions to the one-dimensional {Schr{\"o}dinger}
  equation in terms of {Neumann} series of {Bessel} functions.
\newblock {\em Applied Mathematics and Computation}, 314:173--192, 2017.

\bibitem{KravchenkoPorter2010}
V.~V. Kravchenko and R.~M. Porter.
\newblock Spectral parameter power series for {Sturm-Liouville} problems.
\newblock {\em Mathematical Methods in the Applied Sciences}, 33(4):459--468,
  2010.

\bibitem{KravchenkoTorba_Neumann_2016}
V.~V. {Kravchenko} and S.~M. {Torba}.
\newblock {A Neumann series of Bessel functions representation for solutions of   Sturm-Liouville equations}.
\newblock {\em ArXiv e-prints 1612:08803}, Dec. 2016.

\bibitem{KunitomoIkeda1992}
N.~Kunitomo and M.~Ikeda.
\newblock Pricing options with curved boundaries.
\newblock {\em Mathematical Finance}, 2(4):275--298, 1992.

\bibitem{LadyzhenskayaSU1988}
O.~A. Ladyzhenskaya, V.~A. Solonnikov, and N.~N. Uraltseva.
\newblock {\em Linear and Quasi-Linear Equations of Parabolic Type}, volume~23.
\newblock American Mathematical Society, 1988.

\bibitem{Linetsky2004}
V.~Linetsky.
\newblock The spectral decomposition of the option value.
\newblock {\em International Journal of Theoretical and Applied Finance},
  7(3):337--384, 2004.

\bibitem{LinetskyMendoza-Arriaga2011}
V.~Linetsky and R.~Mendoza-Arriaga.
\newblock Unified credit-equity modeling.
\newblock In T.~R. Bielecki, D.~Brigo, and F.~Patras, editors, {\em Credit Risk
  Frontiers: Subprime Crises, Pricing and Hedging, {CVA}, {MBS}, Ratings, and
  Liquidity}, chapter~18, pages 553--583. Bloomberg Press, New Jersey, 2011.

\bibitem{ArriagaCarrLinetsky2010}
R.~Mendoza-Arriaga, P.~Carr, and V.~Linetsky.
\newblock Time-changed {M}arkov processes in unified credit-equity modeling.
\newblock {\em Mathematical Finance}, 20(4):527--569, 2010.

\bibitem{MijatovicPistorius2013}
A.~Mijatovi\'{c} and M.~Pistorius.
\newblock Continuously monitored barrier options under {M}arkov processes.
\newblock {\em Mathematical Finance}, 23(1):1--38, 2013.

\bibitem{Mikhailov1978}
V.~P. Mikhailov.
\newblock {\em Partial Differential Equations}.
\newblock Mir Publishers, Moscow, 1978.

\bibitem{Nunes2009JFQA}
J.~P. Nunes.
\newblock Pricing {A}merican options under the constant elasticity of variance
  model and subject to bankruptcy.
\newblock {\em Journal of Financial and Quantitative Analysis},
  44(5):1231--1263, 2009.

\bibitem{NunesDiasRuas2017}
J.~P. Nunes, J.~C. Dias, and J.~P. Ruas.
\newblock The early exercise boundary under the jump to default extended {CEV}
  model.
\newblock 2017.
\newblock ISCTE-IUL.

\bibitem{NunesRuasDias2015}
J.~P. Nunes, J.~P. Ruas, and J.~C. Dias.
\newblock Pricing and static hedging of {A}merican-style knock-in options on
  defaultable stocks.
\newblock {\em Journal of Banking and Finance}, 58:343--360, 2015.

\bibitem{Oksendal1995}
B.~{\O}ksendal.
\newblock {\em Stochastic Differential Equations: An Introduction with
  Applications}.
\newblock Springer-Verlag, Berlin Heidelberg, 6th edition, 2003.

\bibitem{Pelsser2000}
A.~Pelsser.
\newblock Pricing double barrier options using {L}aplace transforms.
\newblock {\em Finance and Stochastics}, 4:95--104, 2000.

\bibitem{PinchoverRubinstein2005}
Y.~Pinchover and J.~Rubinstein.
\newblock {\em An Introduction to Partial Differential Equations}.
\newblock Cambridge University Press, United Kingdom, 2005.

\bibitem{PolianinZaitsev1970}
A.~Polianin and V.~Zaitsev.
\newblock {\em Handbook of Ordinary Linear Differential Equations}.
\newblock Faktorial Moskva, 2nd edition, 1970.
\newblock (In Russian).

\bibitem{RogersWilliams1994}
L.~C.~G. Rogers and D.~Williams.
\newblock {\em Diffusions, Markov Processes and Martingales}, volume 1:
  Foundations.
\newblock John Wiley \& Sons, Chichester, England, 2nd edition, 1994.

\bibitem{RuasDiasNunes2013}
J.~P. Ruas, J.~C. Dias, and J.~P. Nunes.
\newblock Pricing and static hedging of {A}merican options under the jump to
  default extended {CEV} model.
\newblock {\em Journal of Banking and Finance}, 37(11):4059--4072, 2013.

\bibitem{Schroder2000}
M.~Schr{\"o}der.
\newblock On the valuation of double-barrier options: Computational aspects.
\newblock {\em Journal of Computational Finance}, 3(4):5--33, 2000.

\bibitem{Shaw1998}
W.~Shaw.
\newblock {\em Modelling Financial Derivatives with {M}athematica}.
\newblock Cambridge University Press, Cambridge, UK, 1998.

\bibitem{Sidenius1998}
J.~Sidenius.
\newblock Double barrier options: Valuation by path counting.
\newblock {\em Journal of Computational Finance}, 1(3):63--79, 1998.

\bibitem{StakgoldHolst2011}
I.~Stakgold and M.~Holst.
\newblock {\em Green's Functions and Boundary Value Problems}.
\newblock Wiley, Hoboken, New Jersey, 3rd edition, 2011.

\bibitem{ZhangZhouZhu2009}
B.~Y. Zhang, H.~Zhou, and H.~Zhu.
\newblock Explaining credit default swap spreads with the equity volatility and
  jump risks of individual firms.
\newblock {\em Review of Financial Studies}, 22(12):5099--5131, 2009.

\end{thebibliography}

\end{document}